\documentclass[aps,pra,twocolumn,epsfig,floats,superscriptaddress]{revtex4}
\usepackage{amsmath}
\usepackage{color}
\usepackage{graphicx}
\usepackage{amsfonts}
\usepackage{mathrsfs}
\usepackage{float}
\usepackage{amsmath}
\usepackage{amssymb}
\usepackage{physics}
\usepackage[utf8x]{inputenc}
\usepackage{times}
\usepackage{hyperref}
\hypersetup{
  colorlinks=true,
  citecolor=blue,
  linkcolor=blue,
  urlcolor=blue}

\DeclareSymbolFont{rmlargesymbols}{OMX}{mdbch}{m}{n}
\DeclareMathSymbol{\rmintop}{\mathop}{rmlargesymbols}{82}

\begin{document}

\title{Impurities in quasi-one-dimensional droplets of binary Bose mixtures}
 
\author{Sudip Sinha}
\affiliation{Indian Institute of Science Education and Research Kolkata, Mohanpur, Nadia 741246, India}
\author{Sayak Biswas}
\affiliation{Physics Department, The Ohio State University, Columbus, Ohio 43210, USA}
\author{L. Santos}
\affiliation{Institut f\"{u}r Theoretische Physik, Leibniz Universit\"{a}t Hannover, 30167 Hannover, Germany}
\author{S. Sinha}
\affiliation{Indian Institute of Science Education and Research Kolkata, Mohanpur, Nadia 741246, India}

\date{\today}

\begin{abstract}
Recently created self-bound quantum droplets of binary Bose mixtures open intriguing possibilities for the study of impurity physics. We show that the properties of impurities embedded in quasi-one-dimensional droplets are determined by the interplay between back-action and quantum fluctuations. Due to such back-action, repulsive impurities may form a metastable quasi-bound state inside the droplet. In contrast, attractive impurities remain bound to the droplet, leading to the hybridization of droplet and impurity excitations, as well as to peculiar scattering resonances. Interestingly, impurity trapping may result solely from the effect of quantum fluctuations. These results may readily be probed experimentally by doping the currently available droplets of binary mixtures.
\end{abstract}

\maketitle
\section{Introduction}
\label{introduction}
A remarkable manifestation of zero point quantum fluctuations has recently been observed in the formation of self bound droplets of ultracold atomic systems. The droplet formation due to beyond mean-field quantum correction, commonly known as Lee-Huang-Yang (LHY) term, has originally been considered in binary mixture of Bose condensates \cite{petrov1} and subsequently demonstrated in experiments \cite{cabrera,semeghini}. Moreover, in recent experiments it has been found that LHY type quantum fluctuations can prevent the collapse of a dipolar condensate leading to the generation of arrays of elongated droplets \cite{pfau1,pfau_prl,pfau2,chomaz}. Usually quantum fluctuations are small in dilute quantum gases, however in certain situations, it can become comparable with the mean-field interactions, and self bound droplets with flat density are produced as a result of competition between them.
The nucleation of such quantum droplets as well their properties and excitations constitute the focus of intense research in recent years \cite{blakie1,santos1,droplet_crystal,blakie2,fort,
Luis_binary_droplet}.
The experimental observation of the crossover from droplet to soliton in quasi-one-dimensional trap \cite{droplet_soliton} has generated impetus to study droplets in lower dimensional geometries \cite{astrakharchik1,edler,petrov2,edmonds}.
Such crossover from droplet to soliton has also been investigated theoretically in quasi-one-dimensional dipolar condensates by tuning the relative strength of attractive dipolar and repulsive short range interactions \cite{edler}.

In a different context, quantum fluctuations become important when mobile impurity atoms in a condensate strongly interact with the phonon modes leading to the formation of heavier quasi-particles known as polarons \cite{Landau,f_polaron_exp1,f_polaron_exp2,
f_polaron_exp3,polaron_effect_expt,b_polaron_exp1,b_polaron_exp2,
b_polaron_critical_exp,Frohlich_original,Demler1,Demler2,
Demler3,bruderer_1}.
Localization and self trapping of polarons is a long standing issue, which has been investigated for dilute condensates \cite{Landau,p_loc_Timmermans,p_loc_blume,bruderer_2}. Interestingly, impurities in a quantum droplet may probe the droplet properties \cite{wenzel,Minardi_2022,Abdullaev_2020}. In this work, we consider the droplet phase of quasi-one-dimensional two-component Bose-Bose mixtures in the presence of impurities, and investigate the formation of their bound states for both repulsive and attractive interactions, a problem that resembles the bosonic version of a quantum dot \cite{Recati_quantum_dot}.  
Incorporating quantum fluctuations, we also study the collective excitations and dynamics of this composite system.

The rest of the paper is organized as follows. In Sec.\ref{model_and_formalism}, we describe the model for a bosonic impurity interacting with a droplet formed by a binary Bose mixture. Section \ref{quasi_bound_state} is devoted to the quasi-bound states of repulsive impurities. In case of attractive impurities, true bound state is formed, for which the collective excitations and intriguing resonance phenomena are discussed in Sec.\ref{attractive_impurity}. We study the quench dynamics for both the attractive as well as repulsive impurities in Sec.\ref{dynamics}. The trapping of impurities solely due to quantum fluctuations is discussed in Sec.\ref{impurity_trapping}. Finally, we summarize our results in Sec.\ref{conclusions}.

\section{Model and formalism}
\label{model_and_formalism}
We consider a two-component Bose mixture with short range interactions in the presence of bosonic impurity atoms in a quasi-one-dimensional (1D) geometry, which is provided by a strong harmonic confinement with frequency $\omega_\perp$ in the transverse direction. Within the single mode approximation, the above mentioned quasi 1D system is described by the Hamiltonian $\hat{\mathcal{H}} = \hat{\mathcal{H}}_{\rm B} + \hat{\mathcal{H}}_{\rm I} + \hat{\mathcal{H}}_{\rm IB}$, with:
\begin{subequations}
%\begin{align}
%\hat{\mathcal{H}}_{\rm B} &=\sum_{i=1,2}  \int \!\! dx\, \hat{\Phi}^\dagger_{i}(x)\!\left[\hat{\mathcal{H}}^{i}_{0}+\frac{g_{i}}{2}\hat{\Phi}^\dagger_{i}(x)\hat{\Phi}_{i}(x)\right]\!\hat{\Phi}_{i}(x)\notag\\
%& \quad\, + g_{12}\!\int\!\! dx\, \hat{\Phi}^\dagger_{1}(x)\hat{\Phi}_{1}(x)\hat{\Phi}^\dagger_{2}(x)\hat{\Phi}_{2}(x)\\
%\hat{\mathcal{H}}_{\rm I} &=\! \int \!\! dx\, \hat{\Psi}^\dagger_{\rm I}(x)\!\left(\!-\frac{\hbar^2}{2m_{\rm I}}\frac{d^2}{dx^2}\right)\!\hat{\Psi}_{\rm I}(x)\label{Hi}\\
%\hat{\mathcal{H}}_{\rm IB} &=\! \sum_{i=1,2}\lambda_{i}\!\!\int \!\! dx\, \hat{\Phi}^\dagger_{i}(x)\hat{\Phi}_{i}(x) \hat{\Psi}^\dagger_{\rm I}(x)\hat{\Psi}_{\rm I}(x) 
%    \label{Hib}
%\end{align} 
\begin{eqnarray}
\hat{\mathcal{H}}_{\rm B} &\!\!=\!\!&\sum_{i=1,2}  \int \!\! dx\, \hat{\Phi}^\dagger_{i}(x)\bigg[\hat{\mathcal{H}}^{i}_{0}+\frac{g_{i}}{2}\hat{\Phi}^\dagger_{i}(x)\hat{\Phi}_{i}(x)\bigg]\hat{\Phi}_{i}(x)\notag\\
&&+ g_{12}\!\int\!\! dx\, \hat{\Phi}^\dagger_{1}(x)\hat{\Phi}_{1}(x)\hat{\Phi}^\dagger_{2}(x)\hat{\Phi}_{2}(x)\\
\hat{\mathcal{H}}_{\rm I} &\!\!=\!\!& \int \!\! dx\, \hat{\Psi}^\dagger_{\rm I}(x)\!\left(\!-\frac{\hbar^2}{2m_{\rm I}}\frac{d^2}{dx^2}\right)\!\hat{\Psi}_{\rm I}(x)\label{Hi}\\
\hat{\mathcal{H}}_{\rm IB} &\!\!=\!\!& \sum_{i=1,2}\lambda_{i}\!\!\int \!\! dx\, \hat{\Phi}^\dagger_{i}(x)\hat{\Phi}_{i}(x) \hat{\Psi}^\dagger_{\rm I}(x)\hat{\Psi}_{\rm I}(x) 
    \label{Hib}
\end{eqnarray}   
\end{subequations}
where $\hat{\mathcal{H}}^{i}_{0}\!\!=\!\!-\frac{\hbar^2}{2m_{i}}\frac{d^2}{dx^2}\!-\!\mu_{i}$ with $m_{i}$ ($\mu_{i}$) being the mass (chemical potential) of the $i^{\rm th}$ species ($i=1,2$). The Hamiltonians $\hat{\mathcal{H}}_{\rm B}$ and $\hat{\mathcal{H}}_{\rm I}$ represent the two-component mixture and impurity atoms (with mass $m_{\rm I}$), described by the bosonic field operators $\hat{\Phi}_{i}(x)$ and $\hat{\Psi}_{\rm I}(x)$, respectively. In terms of the s-wave scattering length $a_{i}$ ($a_{12}$), the inter (intra)-species interaction strength is given by $g_{i}=2a_{i}\hbar^2/m_{i} l^2_{\perp}$ ($g_{12}=a_{12}\hbar^2/m_{\rm r}l^2_{\perp}$), where $l_{\perp}=(\hbar/m_{\rm r}\omega_{\perp})^{1/2}$ is the transverse confinement length and the reduced mass $m_{\rm r} = m_{1}m_{2}/(m_{1}+m_{2})$.
 The Hamiltonian $\hat{\mathcal{H}}_{\rm IB}$ describes the interaction between the bosonic species and impurity atoms with strength $\lambda_{i}=a^{\rm I}_{i}\hbar^2/m^{\rm I}_{{\rm r}i}l^2_{\perp}$, where $a^{\rm I}_{i}$ and $m^{\rm I}_{{\rm r}i} = m_{i}m_{\rm I}/(m_{i}+m_{\rm I})$ denote the associated s-wave scattering length and reduced mass, respectively. We consider a low concentration of impurity atoms $n_{\rm I}\ll n_{i}$, where $n_{\rm I}$ and $n_{i}$ correspond to the density of impurity atoms and bosonic species, respectively. 
We consider $a_{i}/l_{\perp}\!\!\ll\!\!1$, $a^{\rm I}_{i}/l_{\perp}\!\!\ll\!\!1$ and $|\mu_{i}|/\hbar\omega_{\perp}\!\!\ll\!\!1$ which ensure the validity of the single mode approximation in such quasi 1D regime. In the rest of the paper, as well as in all the figures, we set $\hbar\!=\!1$, and scale the energies, time and lengths by $\omega_{\perp}$, $1/\omega_{\perp}$, and $l_{\perp}$, respectively.

For the quasi 1D bosons described by $\hat{\mathcal{H}}_{\rm B}$, the weakly interacting regime for uniform Bose gases (with density $n_{i}$) can be achieved when the dimensionless parameter $\gamma_{i}=2a_{i}/n_{i}l^2_{\perp}\ll 1$ \cite{Shlyapnikov,Y_Castin}. In general, within the Bogoliubov approximation, the field operators $\hat{\Phi}_{i}(x)$ can be decomposed as, 
\begin{eqnarray}
\hat{\Phi}_{i}(x) = \phi_{i}(x) + \sum_{j,\nu\neq 0} u_{j,\nu}(x) \hat{\alpha}_{j,\nu}-v^{*}_{j,\nu}(x)\hat{\alpha}^\dagger_{j,\nu} , 
\end{eqnarray}
where $j=1,2$; $\phi_{i}(x)$ represents the wavefunction of the quasi condensate and the quantum fluctuations are described in terms of annihilation (creation) operator $\hat{\alpha}_{j,\nu} (\hat{\alpha}^\dagger_{j,\nu})$ of Bogoliubov excitations with energy $\mathcal{E}_{\pm,\nu}$ \cite{stringari_pitaevskii_book}. 
Up to the quadratic fluctuations, the total Hamiltonian can be written as,
\begin{eqnarray}
\hat{\mathcal{H}} &\!\!=\!\!& \sum_{i=1,2}\bigg[\int \!\! dx\, \phi^{*}_{i}(x)\!\left(-\frac{\hbar^2}{2m_{i}}\frac{d^2}{dx^2}\!-\!\mu_{i}\!+\!\frac{g_{i}}{2}n_{i}(x)\right)\!\phi_{i}(x)\notag\\
&&+\lambda_{i}\!\int \!\! dx\, n_{i}(x)\hat{n}_{\rm I}(x)\bigg] + E_{\text{\tiny LHY}} + \hat{\mathcal{H}}_{\rm I} + \hat{\mathcal{H}}_{\rm F}\notag\\
&&+\sum_{\nu \neq 0}\left( \mathcal{E}_{+,\nu}\,\hat{\alpha}^\dagger_{1,\nu}\,\hat{\alpha}_{1,\nu}+\mathcal{E}_{-,\nu}\,\hat{\alpha}^\dagger_{2,\nu}\,\hat{\alpha}_{2,\nu}\right)
\end{eqnarray}
%\begin{align}
%&\hat{\mathcal{H}} \!= \!\!\!\!\sum_{i=1,2}\!\!\left[\int \!\! dx\, \phi^{*}_{i}(x)\!\!\left(\!\hat{\mathcal{H}}^{i}_{0}\!+\!\frac{g_{i}}{2}n_{i}(x)\!\right)\!\!\phi_{i}(x)\!+\!\lambda_{i}\!\!\int \!\! dx\, n_{i}(x)\hat{n}_{\rm I}(x)\!\right]\notag\\
%&+\!E_{\text{\tiny LHY}}\!+\!\hat{\mathcal{H}}_{\rm I}\!+\!\hat{\mathcal{H}}_{\rm F}\!+\!\!\!\sum_{\nu \neq 0}( \mathcal{E}_{+,\nu}\hat{\alpha}^\dagger_{1,\nu}\hat{\alpha}_{1,\nu}\!+\mathcal{E}_{-,\nu}\hat{\alpha}^\dagger_{2,\nu}\hat{\alpha}_{2,\nu})
%\end{align}
where $n_{i}(x)\!=\!|\phi_{i}(x)|^2$ and $\hat{n}_{\rm I}(x)\!=\!\hat{\Psi}^{\dagger}_{\rm I}(x)\hat{\Psi}_{\rm I}(x)$. The Hamiltonian $\hat{\mathcal{H}}_{\rm F}$ describes the phonon impurity interaction in the weak coupling regime, known as the Fr\"ohlich term \cite{Demler1,Demler2,Demler3}. For homogeneous Bose gases with density $n_{i}$, the Bogoliubov modes can be written in terms of plane wave basis ($\nu \equiv k$) and the Fr\"ohlich term $\hat{\mathcal{H}}_{\rm F}$ reduces to,
\begin{eqnarray}
\hat{\mathcal{H}}_{\rm F} = \sum_{i, k\neq 0}\lambda_{i}\,\sqrt{n_{i}}\,\hat{\rho}_{{\rm I}, k}\,\left(\hat{a}_{i,k}+\hat{a}^{\dagger}_{i,-k}\right)
%\hat{\mathcal{H}}_{\rm F} =&\sum_{i; k\neq 0}\lambda_{i}\sqrt{n_i}\,\hat{\rho}_{{\rm I},k}\,\frac{k^2}{m_{i}}\,\Big[\mathcal{N}^{(i)}_{+,k}\left(\hat{\alpha}_{1,k}+\hat{\alpha}^\dagger_{1,-k}\right)\notag\\
%&+\mathcal{N}^{(i)}_{-,k}\left(\hat{\alpha}_{2,k}+\hat{\alpha}^\dagger_{2,-k}\right)\Big]
\end{eqnarray}
where $\hat{\rho}_{{\rm I},k}=\int\!dx \, \hat{n}_{\rm I}(x)e^{\imath k x}/\sqrt{L}$ and $\hat{a}_{i,k} (\hat{a}^\dagger_{i,k})$ denotes the  annihilation (creation) operator for bosonic species which can be written in terms of the Bogoliubov operators $\hat{\alpha}_{i,k}$ (see Appendix \ref{quantum_correction_binary_mixture_appendix} for details). The Lee-Huang-Yang (LHY) correction \cite{petrov1,cabrera,semeghini} to the ground state energy due to the zero-point fluctuations of the Bogoliubov modes of the binary mixture is given by,
\begin{eqnarray}
\frac{E_{\text{\tiny LHY}}}{L} = \frac{1}{2}\!\int\! \frac{dk}{2\pi}\left[\mathcal{E}_{+,k}+\mathcal{E}_{-,k}-\sum_{i=1,2}\left(\frac{k^2}{2m_{i}}+g_{i}n_{i}\right) \right]
\label{LHY_energy_correction}
\end{eqnarray} 
where the Bogoliubov excitation branches $\mathcal{E}_{\pm,k}$ are given by,
\begin{eqnarray}
\mathcal{E}_{\pm,k} = \frac{\mathcal{E}^2_{1,k}+\mathcal{E}^2_{2,k}}{2}\pm\sqrt{\frac{(\mathcal{E}^2_{1,k}-\mathcal{E}^2_{2,k})^2}{4}+g^2_{12}\frac{n_{1}n_{2}}{m_{1}m_{2}}k^4}
\label{two_comp_excitation_branches}
\end{eqnarray}
with $\mathcal{E}_{i,k} = [\frac{k^2}{2m_{i}}(\frac{k^2}{2m_{i}}+2g_{i}n_{i})]^{1/2}$. 
For simplicity, we consider equal density profiles $n_{1}(x)\!=\!n_{2}(x)\!=\!n(x)$ with same intra-species coupling $g_{1}\!=\!g_{2}\!=\!g$ and masses $m_{1}\simeq m_{2}=m$, and hence the binary mixture reduces to an effective single component system with density $n(x)=|\phi(x)|^2$ \cite{petrov2,astrakharchik1}. In this case, the correction to the chemical potential due to the LHY term reduces to,
\begin{eqnarray}
\Delta \mu_{\text{\tiny LHY}} = -\frac{\sqrt{g^3 n m}}{2 \pi}\sum_{\pm}\left(1\pm\sqrt{\eta}\right)^{3/2}
\end{eqnarray}
where $\eta=g^2_{12}/g^2$. 
Here we consider purely 1D LHY correction, which may however undergo a dimensional crossover for sufficiently large $n_{i}a_{i}$ \cite{dimensional_crossover}. 
By treating $\hat{\mathcal{H}}_{\rm F}$ perturbatively, the second-order correction to the ground state energy due to $N_{\rm I}$ impurity particles can be computed (see Appendix \ref{quantum_correction_due_to_impurities_appendix} for details) and the corresponding correction to the impurity chemical potential is given by,
\begin{eqnarray}
\Delta \mu_{\rm I}=-\frac{\sqrt{(n/g)m}}{2 \pi}\Lambda(\alpha)\!\!\sum_{\pm}\!\frac{(\lambda_1\pm\lambda_2)^{2}\!\left(1\mp\sqrt{\eta}\right)^{1/2}}{\sqrt{1-\eta}}
\label{delmu_I}
\end{eqnarray}
where $\Lambda(\alpha)=\sin^{-1}(\sqrt{1-\alpha^2})/\sqrt{1-\alpha^2}$ and $\alpha = m/m_{\rm I}$. For a dilute gas of non-interacting impurity atoms, we can neglect the higher-order corrections $\Delta \mu^{(i)}_{\rm I}$ arising from them, as $\Delta \mu^{(i)}_{\rm I} \ll \Delta \mu_{\text{\tiny LHY}}$. Treating the above corrections within the local density approximation (LDA), we can write the effective Gross-Pitaevskii equations (GPE) for the single component condensate and impurities with wavefunctions $\phi(x,t)$ and $\psi_{\rm I}(x,t)$ respectively (see also Appendix \ref{appendix_local_density_approx}),
\begin{subequations}
\begin{eqnarray}
i\partial_{t}\phi &\!\!=\!\!&\left(-\frac{1}{2m}\frac{\partial^2}{\partial x^2}+\delta gn+\Delta \mu_{\text{\tiny LHY}}\!+\!\frac{\tilde{\lambda}}{2}n_{\rm I}\right)\phi
\label{effective_GPE_1} \\
i\partial_{t}\psi_{\rm I} &\!\!=\!\!&\Bigg(-\frac{\alpha}{2m}\frac{\partial^2}{\partial x^2} + \tilde{\lambda}n+\Delta \mu_{\rm I}\Bigg)\psi_{\rm I}
\label{effective_GPE_2}
\end{eqnarray}
\label{effective_GPE}
\end{subequations}\\
where $\delta g = g+g_{12}$ and $\tilde{\lambda} = \lambda_{1}+\lambda_{2}$.  
For a quasi 1D Bose gas, the LHY correction is attractive \cite{astrakharchik1,petrov2}, whereas the mean-field contribution is repulsive for $0<\delta g\ll g$. By tuning $\delta g/g$ sufficiently close to zero, the mean-field part can become comparable with the LHY correction, and the competition between them gives rise to a self bound droplet with flat density for a sufficiently large number of bosons, which is the main focus of the present work.

\begin{figure}
    \centering
    \includegraphics[width=1.01\columnwidth]{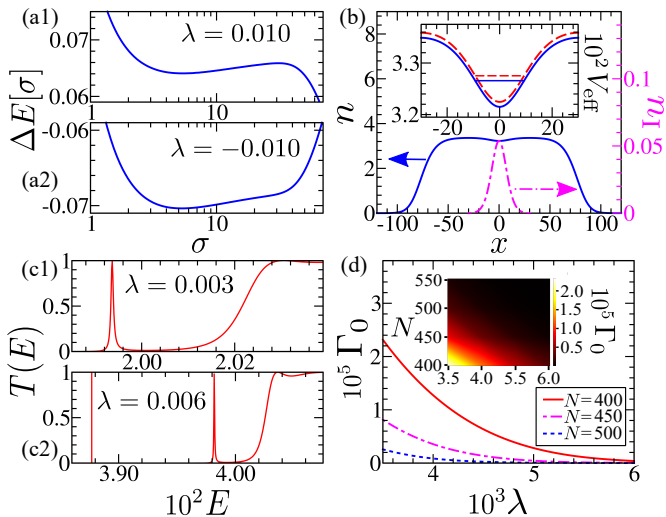}
    \caption{{\it Droplet-embedded impurities for $\lambda_{1,2}=\lambda$}: Variation of the additional energy cost $\Delta E[\sigma]$ with the impurity width $\sigma$ in case for (a1) repulsive $(\lambda>0)$ and (a2) attractive $(\lambda<0)$ coupling for $N=500$. (b) Density profiles of the droplet $n$ (left axis) and impurity $n_{\rm I}$ (right axis) for a repulsive coupling strength $\lambda = 0.005$. The inset shows the effective potential $V_{\rm eff}(x)$ zoomed at the peak of the droplet with (solid blue line) and without (red dashed line) quantum correction, with the horizontal lines (blue and red, respectively) denoting the lowest energy of the quasi-bound states. (c1,c2) Transmission coefficients $T(E)$ for different values of $\lambda$. 
    %The sharp resonance peaks correspond to the energy of the quasi-bound states. 
    (d) Variation of the scattering resonance width $\Gamma_{0}$ with $\lambda$ for different values of $N$. The inset contains the color scaled plot of $\Gamma_{0}$ in the parameter space $(N, \lambda)$. In all the figures we employ: $g= 0.1$, $g_{12} = -0.95g$, and $\alpha=0.1$.}
    \label{fig:1}
\end{figure}

\section{Quasi-bound state for repulsive impurities}
\label{quasi_bound_state}
We investigate in the following the properties of impurities embedded in self bound droplet of a binary mixture, using the effective GPE in Eq.~\eqref{effective_GPE}.
We consider first the case of a repulsive impurity. Although the formation of polaronic bound states \cite{Landau} of repulsive impurities in homogeneous condensates has been considered \cite{p_loc_Timmermans,p_loc_blume,bruderer_2}, it remains unclear in the case of a self bound droplet with finite size, since the impurity atoms can separate out from the droplet due to repulsion. 
Here, we investigate the possible formation of bound states of repulsive impurities using a Gaussian variational wavefunction localized at the center of the droplet, 
\begin{eqnarray}
\psi(x)=(1/\sigma\sqrt{\pi})^\frac{1}{2}e^{-x^2/2\sigma^2} 
\end{eqnarray}
where the width $\sigma$ is considered as the variational parameter. We numerically compute the stationary state of the droplet using Eq.~\eqref{effective_GPE_1}, which in turn yields the effective potential for the impurity [see Eq.~\eqref{effective_GPE_2}]. 
Following this, we compute the additional energy cost $\Delta E[\sigma]$ of these impurity-droplet configurations due to the presence of impurities, as a function of $\sigma$.
Interestingly, the appearance of a local minimum in $\Delta E[\sigma]$ for small $\sigma$ indicates a metastable bound state of impurities at the center of the droplet, as depicted in Fig.\ref{fig:1}(a1). Such metastable state is protected by an energy barrier, above which, $\Delta E[\sigma]$ decreases with increasing $\sigma$, exhibiting unbinding of the impurity. Intuitively, the impurity in such a state would remain localized as long as fluctuations are not strong enough to overcome such energy barrier. As shown in Fig.\ref{fig:1}(b), a dip in the density profile at the center of the droplet gives rise to a well-like structure in the effective potential experienced by the impurities,
\begin{eqnarray}
V_{\text{eff}} = \tilde{\lambda}n(x)+\Delta\mu_{\rm I}
\end{eqnarray}  
The stationary states of this potential well correspond to the quasi-bound states of the impurities, which can decay into the continuum of states outside the droplet. The ground state wavefunction of the impurity can be obtained by solving Eq.~\eqref{effective_GPE}(a,b) in a self consistent manner considering only the local effective potential at the center of droplet.  
Numerically, a rapid convergence to the quasi-bound state in the local  effective potential is possible, as long as it is well separated from the scattering continuum. We also verify that such impurity quasi-bound state remain a stationary solution for sufficiently long time by evolving Eq.~\eqref{effective_GPE}(a,b).
 
An important characteristic of such metastable quasi-bound state is its lifetime, which can be computed from the scattering of impurity particles in the presence of the effective potential $V_{\rm eff}$ obtained self consistently as mentioned above. 
The transmission coefficient $T(E)$ exhibits a series of shape resonances due to the quasi-bound states of $V_{\rm eff}$, as shown in Fig.\ref{fig:1}(c1,c2). The resonance peak with width $\Gamma_{0}$ at the lowest energy corresponds to the quasi-bound impurity state with lifetime $\tau=1/\Gamma_0$. The stability of such an impurity state in terms of its lifetime is shown in Fig.\ref{fig:1}(d).  
Increasing the repulsive strength $\lambda$ of the impurities makes the effective potential well $V_{\rm eff}$ deeper, supporting the formation of quasi-bound states. 
The stability is further favored when the number of bosons $N$ grows, due to the larger droplet size, which leads to an enhanced separation of the quasi-bound states from the continuum.
Dynamical unbinding of such quasi-bound impurity can occur by squeezing of the droplet under a sufficiently strong quench in the inter-species interaction strength, which we discuss later in Sec.\ref{dynamics}.

\section{Attractive impurity and collective modes}
\label{attractive_impurity}
Unlike the repulsive case, attractive impurities form true bound states with the droplet. The energy cost $\Delta E[\sigma]$ shows a global minimum [see Fig.\ref{fig:1}(a2)], which corresponds to the localized ground state of the effective potential $V_{\rm eff}$ with energy $\mu_{\rm I}<0$. 
The stationary states of $V_{\rm eff}$ obtained from Eq.~\eqref{effective_GPE_2} correspond to the impurity excitation energies $\epsilon_{\rm I}$ with respect to the ground state $\mu_{\rm I}$, that are bound for $\epsilon_{\rm I} <-\mu_{\rm I}$, and above which, the scattering continuum exists.
On the other hand, the self bound droplet with $\mu<0$ has discrete collective excitations below $-\mu$, above which they merge with the scattering continuum.
Beyond the above mentioned single particle description, the impurity excitations couple to the droplet fluctuations, and the collective excitations $\epsilon$ of the full system can be described by the effective Bogoliubov equations, obtained from linearization of Eq.~\eqref{effective_GPE}(a,b),
\begin{small}
\begin{eqnarray}
\left(\begin{array}{cccc}
\mathcal{L}_{0}+X & X & Y & Y  \\
-X & -\mathcal{L}_{0}-X  & -Y & -Y  \\
\widetilde{Y} & \widetilde{Y} & \mathcal{L}_{\rm I} & 0\\
-\widetilde{Y} & -\widetilde{Y} & 0 & -\mathcal{L}_{\rm I}\\
\end{array}\right)
\left(\!\begin{array}{c}
u^{(1)}\\
v^{(1)}\\
u^{(2)}\\
v^{(2)}\\
\end{array}\!\right) 
=\epsilon\!\left(\!\begin{array}{c}
u^{(1)}\\
v^{(1)}\\
u^{(2)}\\
v^{(2)}\\
\end{array}\!\right)
\label{Bogoliubov_equation}
\end{eqnarray}
\end{small}
where the different matrix elements are given by,
%where $\mathcal{L}_0\!\!=\!\!-\frac{1}{2m}\frac{d^2}{dx^2}\!+\!
%\delta g n(x)\!+\!\tilde{\lambda}n_{\rm I}(x)/2\!-\!\mu\!+\!\Delta \mu_{\text{\tiny LHY}}$, $\mathcal{L}_{\rm I}\!=\!-\frac{\alpha}{2m}\frac{d^2}{dx^2} + \tilde{\lambda} n(x)-\mu_{\rm I}+\Delta \mu_{\rm I}$, $X\!=\!\delta g n(x)+n(x)\frac{\partial \Delta \mu_{\text{\tiny LHY}}}{\partial n(x)}$, $Y\!=\!\tilde{\lambda}\sqrt{n(x)n_{\rm I}(x)}/2$, and $\widetilde{Y}\!=2Y+\frac{\partial\Delta\mu_{\rm I}}{\partial n(x)}\sqrt{n(x)n_{\rm I}(x)}$.
\begin{subequations}
\begin{align}
\mathcal{L}_0 &=-\frac{1}{2m}\frac{d^2}{dx^2}\!+\!\delta g n(x)\!+\!\tilde{\lambda}n_{\rm I}(x)/2\!-\!\mu\!+\!\Delta \mu_{\text{\tiny LHY}} \\
\mathcal{L}_{\rm I}&=-\frac{\alpha}{2m}\frac{d^2}{dx^2} + \tilde{\lambda} n(x)-\mu_{\rm I}+\Delta \mu_{\rm I} \\
X&=\delta g n(x)+n(x)\frac{\partial \Delta \mu_{\text{\tiny LHY}}}{\partial n(x)} \\
Y&=\tilde{\lambda}\sqrt{n(x)n_{\rm I}(x)}/2 \\
\widetilde{Y}&=\tilde{\lambda}\sqrt{n(x)n_{\rm I}(x)}+\frac{\partial\Delta\mu_{\rm I}}{\partial n(x)}\sqrt{n(x)n_{\rm I}(x)}
\end{align}
\end{subequations} 
The droplet and impurity amplitudes are given by $(u^{(1)},v^{(1)})$ and $(u^{(2)},v^{(2)})$ respectively, which satisfy the normalization condition,
\begin{eqnarray} 
\int\!dx \,\left(\,|u^{(1)}|^2-|v^{(1)}|^2+|u^{(2)}|^2-|v^{(2)}|^2\right)=1.
\end{eqnarray}

\begin{figure}
    \centering
    \includegraphics[width=1.02\columnwidth]{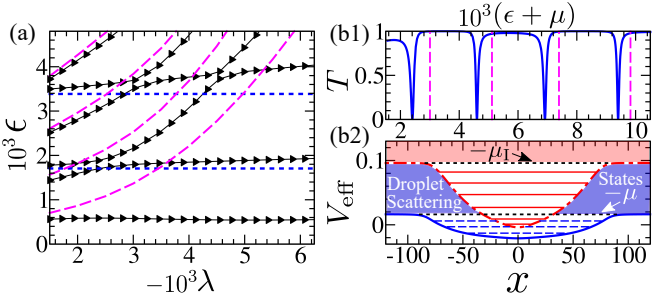}
    \caption{ {\it Formation of bound impurity states in a droplet for attractive interaction $\lambda_{1,2}=-\lambda$}: (a) Variation of the excitation energy $\epsilon$ of the Bogoliubov modes with $-\lambda$. The black triangles represent the excitation modes obtained from solving Eq.~\eqref{Bogoliubov_equation}. The dotted blue (dashed red) line corresponds to droplet (impurity) modes in the decoupled approximation. (b1) Variation of the transmission coefficient $T$ with $\epsilon > -\mu$ for $\lambda=-0.006$. Vertical dashed lines represent the single-particle impurity energy levels $\epsilon_{\rm I}$. (b2) Schematic of single-particle effective potential experienced by the impurity (dashed dotted line) and the droplet (solid line). Single-particle impurity (droplet) energy levels are marked by horizontal solid (dashed) lines, and the scattering states are denoted by shaded regions. The chemical potential values are denoted by horizontal dotted lines marked by arrows. In all figures we employ: $g= 0.1$, $g_{12} =-0.95g$, $\alpha = 0.4$, $N=400$, and $N_{\rm I}=5$.}
    \label{fig:2}
\end{figure}

The excitation spectrum exhibits various interesting features. For $\mu_{\rm I} < \mu$, the bound modes of the droplet (below $-\mu$) can hybridize with the bound impurity modes, giving rise to a new  set of bound excitations. 
Based on the relative strength of the Bogoliubov amplitudes, these can be classified as being droplet-like for $\int\!dx\,(|u^{(1)}|^2-|v^{(1)}|^2)>\int\!dx\, (|u^{(2)}|^2-|v^{(2)}|^2)$ or impurity-like for $\int\!dx \,(|u^{(1)}|^2-|v^{(1)}|^2)<\int\!dx \,(|u^{(2)}|^2-|v^{(2)}|^2)$. In Fig.\ref{fig:2}(a), the impurity-like excitations are compared with the single-particle impurity states obtained from the effective potential $V_{\rm eff}$. Interestingly, within the range $-\mu<\epsilon<-\mu_{\rm I}$, the scattering continuum of droplet excitations coexist with the bound modes of the impurities, which results in an intriguing resonance phenomena in the scattering with impurities.
To study the scattering of the droplet modes with the impurities, we numerically solve Eq.~\eqref{Bogoliubov_equation} following the method outlined in Ref.\cite{DVR}, considering the asymptotically free droplet excitations with $u^{(1)} \sim e^{\pm\imath kx}$ and the amplitude of other channels decaying to zero. We compute the transmission coefficient $T$, which vanishes when the excitation $\epsilon$ becomes close to the single-particle state of the impurities, exhibiting scattering resonances, as shown in Fig.\ref{fig:2}(b).

\section{Quench dynamics of impurity}
\label{dynamics}
\begin{figure}[H]
    \centering
   \includegraphics[width=1.05\columnwidth]{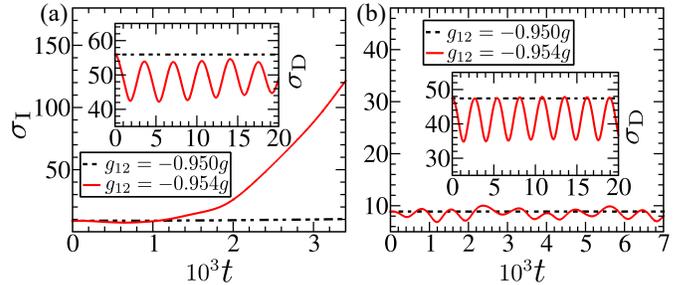}
    \caption{{\it Quench dynamics of impurity:} Contrast in the time evolution of width $\sigma_{\rm I}$ of an initial impurity state for $\lambda_{1,2}=\lambda$ by quenching $g_{12}$ for (a) repulsive $(\lambda=0.006)$ and (b) attractive $(\lambda=-0.006)$ cases. The insets show the time evolution of width $\sigma_{\rm D}$ of the respective droplet states. The initial state in both the cases is prepared for $g=0.1$, $g_{12}=-0.95g$, $\alpha=0.4$, $N=400$, and $N_{\rm I}=5$.}
    \label{fig:3}
\end{figure}

To this end, we discuss the coupled dynamics of impurity and droplet  under a sudden change of the inter-species coupling strength $g_{12}$, affecting the droplet size sensitively. First, we investigate the fate of a sufficiently stable quasi-bound state of a repulsive impurity as $g_{12}$ is increased abruptly, generating a compression of the droplet. Under such perturbation, the droplet exhibits  a breathing oscillation, that leads to unbinding of the impurity from the droplet depending on the strength of the quench, which is manifested as spreading of the impurity wavefunction, as shown in Fig.\ref{fig:3}(a). On the other hand, an attractive impurity remains bound to the droplet and exhibits breathing oscillations, as $g_{12}$ is changed [see Fig.\ref{fig:3}(b)]. Note that, we have not taken into account the three body losses, which may play an important role in the damping of such oscillations.

\section{Impurity trapping due to LHY potential}
\label{impurity_trapping}
\begin{figure}
    \centering
    \includegraphics[width=1.01\columnwidth]{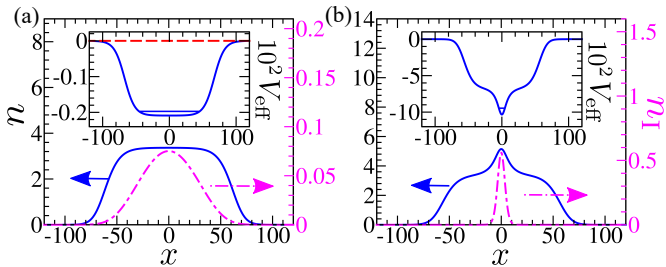}
    \caption{{\it Impurity trapping due to LHY correction}: Comparison of the density profiles of the droplet $n$ (left axis) and impurity $n_{\rm I}$ (right axis) for (a) $\lambda_{1}=-\lambda_{2}=0.010$ and (b) $\lambda_{1}=\lambda_{2}=-0.010$. The inset shows the effective potential $V_{\rm eff}$ with (solid blue line) and without (red dashed line) quantum correction, where the horizontal solid blue lines denote the ground state energy of the impurity bound state. Parameters chosen: $g=0.1$, $g_{12}=-0.95g$, $\alpha=0.4$, $N = 400$, and $N_{\rm I}=5$.}
    \label{fig:4}
\end{figure}

We consider at this point the situation when the mean-field effective potential experienced by the impurity becomes vanishingly small, and the impurity atoms interact with two components of the droplet in an opposite manner (demanding $\lambda_{1}=-\lambda_{2}$). In this situation, the mean-field potential itself is unable to bind the impurity, however the LHY quantum correction gives rise to an effective attractive potential well, 
\begin{eqnarray}
V_{\rm eff} = \Delta \mu_{\rm I}<0,
\end{eqnarray}
which can lead to the formation of an impurity bound state. In contrast to the attractive impurity interaction $\lambda_{1,2}=-\lambda$ with very localized impurity bound states [see Fig.\ref{fig:4}(b)], in this case, the effective potential takes the shape of a flat well, as a result, the wavefunction of the impurity spreads and its width becomes comparable with the droplet size, as shown in Fig.\ref{fig:4}(a). This corresponds to a unique situation of trapping impurity atoms purely due to the quantum fluctuations.

\section{Conclusions}
\label{conclusions}
 We have investigated the intriguing physics of impurities in one-dimensional quantum droplets of binary mixtures. For both repulsive and attractive impurities, such composite system was analyzed within the framework of coupled Gross-Pitaevskii equations, incorporating quantum fluctuations in a systematic manner. On one hand, the repulsive interaction between impurity and bosons lead to the formation of metastable quasi-bound states with finite lifetime, as a result of the impurity back-action on the droplet. On the other hand, for the attractive impurity-boson interactions, a true bound state is formed, for which we obtain the collective excitations, exhibiting hybridization of impurity and droplet modes. Furthermore, an intriguing resonance phenomena is observed when the free particle excitations of the droplet coincide with bound excitations of the impurity.  Moreover, we identify a regime where a unique situation can arise when the impurity can form bound states solely due to the quantum fluctuations. This impurity trapping phenomena can also serve as a probe for quantum fluctuations. Finally, we study the quench dynamics, which leads to the unbinding of the repulsive impurity from the droplet. On the contrary, the attractive impurities remain bound to the droplet, exhibiting breathing oscillations. Experimental progress on mixtures of Bose gases shows promise to realize such (quasi) bound states of impurities in a quantum droplet, which can provide a unique opportunity to study few body correlations as well as collective properties of the droplet.

\section*{ACKNOWLEDGMENTS}
L. S. acknowledges the support of the Deutsche Forschungsgemeinschaft (DFG, German Research Foundation) under Germany’s Excellence Strategy – EXC-2123 Quantum-Frontiers – 390837967.

\appendix
\section{Quantum corrections within the Bogoliubov approach}
\label{quantum_correction_bogoliubov_appendix}
Here, we derive the quantum corrections arising from the interaction between the two components of the binary mixture as well as from the interaction of the impurities with the binary mixture. 

For the binary Bose-Bose mixture, we consider homogeneous condensates with densities $n_{i}$ ($i=1,2$), and expand the corresponding bosonic field operators in the plane wave basis,
\begin{eqnarray}
\hat{\Phi}_{i}(x) = \sqrt{n_{i}}+\frac{1}{\sqrt{L}}\sum_{k\neq0}e^{\imath kx}\hat{a}_{i,k}
\end{eqnarray} 
where $\hat{a}_{i,k}$ ($i=1,2$) denotes the annihilation operator for species 1(2). The Hamiltonian in Eq.~(1) of the main text can be expanded up to quadratic order as follows,
\begin{eqnarray}
\hat{\mathcal{H}} &\!\!=\!\!& E_{\text{\tiny MF}} + \hat{\mathcal{H}}_{c} + \hat{\mathcal{H}}_{\rm I}+ \sum_{i=1,2}\lambda_{i}\!\int\!dx\,\hat{n}_{\rm I}(x) n_{i} \notag\\
&&+ \sum_{i, k\neq 0}\lambda_{i}\sqrt{n_{i}}\,\hat{\rho}_{\text{\tiny I},k}(\hat{a}_{i,k}+\hat{a}^\dagger_{i,-k})
\label{background_expansion}
\end{eqnarray}
where $E_{\text{\tiny MF}}=L(g_{1}n^2_{1}/2+g_{2}n^2_{2}/2+g_{12}n_{1}n_{2})$ is the mean-field energy and $\hat{\mathcal{H}}_{c}$ represents the fluctuations up to quadratic order for the binary Bose gas. The second last term describes the interaction between the impurities and the binary mixture of bosons, where $\hat{n}_{\rm I}(x) = \hat{\Psi}^\dagger_{\rm I}(x)\hat{\Psi}_{\rm I}(x)$. The last term is called Fr\"ohlich Hamiltonian, which describes the impurity-phonon interaction, where $\hat{\rho}_{{\rm I},k}=\int\!dx\, \hat{n}_{\rm I}(x)e^{\imath k x}/\sqrt{L}$.
%The last two terms are called Fr\"ohlich terms with $\hat{\rho}_{\text{\tiny I},k}=\frac{1}{\sqrt{L}}\int dx\, \hat{n}_{\rm I}(x)e^{\imath k x}$ and $\hat{n}_{\rm I}(x) = \hat{\Psi}^\dagger_{\rm I}(x)\hat{\Psi}_{\rm I}(x)$, which represent the interaction between the impurities and the binary mixture of bosons.\\

\subsection{Lee-Huang-Yang (LHY) correction for a binary mixture of Bose gases}
\label{quantum_correction_binary_mixture_appendix}
In the following, we derive the quantum correction a.k.a the ``Lee-Huang-Yang (LHY)" correction due to the interaction between the two components of the binary Bose gas in 1D. We can write the operators $\hat{a}_{i,k}$ in terms of the Bogoliubov operators $\hat{\alpha}_{i,k}$ by using the Inverse Bogoliubov transformation, which are given as following,
\begin{widetext}
\begin{align}
\hat{a}_{i,k} &= \mathcal{N}^{(i)}_{+,k}\!\left[ \left(\!\mathcal{E}_{+,k}+\frac{\hbar^2k^2}{2m_{i}}\right)\!\hat{\alpha}_{1,k}\!-\!\left(\mathcal{E}_{+,k}-\frac{\hbar^2k^2}{2m_{i}}\right)\!\hat{\alpha}^\dagger_{1,-k}\right]+ \mathcal{N}^{(i)}_{-,k}\!\left[\left(\mathcal{E}_{-,k}+\frac{\hbar^2k^2}{2m_{i}}\right)\!\hat{\alpha}_{2,k}\!-\!\left(\mathcal{E}_{-,k}-\frac{\hbar^2k^2}{2m_{i}}\right)\!\hat{\alpha}^\dagger_{2,-k} \right]
\label{inv_bog_transformation}
\end{align}
where $\mathcal{E}_{\pm,k}$ denote the excitation spectrum for the two-component Bose gas \cite{petrov1} given in Eq.~\eqref{two_comp_excitation_branches}, and the normalization constants $\mathcal{N}^{(i)}_{\pm,k}$ are given by, 
\begin{eqnarray}
\mathcal{N}^{(i)}_{\pm,k}=\left[\frac{1}{4\mathcal{E}_{\pm,k}}\left(\frac{2m_{i}}{\hbar^2k^2}\right)\frac{(\mathcal{E}^2_{\pm,k}-\mathcal{E}^2_{\tilde{i},k})}{(\mathcal{E}^2_{\pm,k}-\mathcal{E}^2_{\mp,k})}\right]^{1/2}
\end{eqnarray}
with $i\neq\tilde{i}$. After doing the Bogoliubov transformation for the two-component Bose gas, the diagonalized part is given by,
\begin{eqnarray}
\hat{\mathcal{H}}^{\text{\tiny D}}_{c} = \sum_{k\neq 0} \left( \mathcal{E}_{+,k}\hat{\alpha}^\dagger_{1,k} \hat{\alpha}_{1,k} + \mathcal{E}_{-,k}\hat{\alpha}^\dagger_{2,k} \hat{\alpha}_{2,k}\right) + E_{\text{\tiny LHY}}
\label{2_comp_quantum_correction}
\end{eqnarray}
The Lee-Huang-Yang (LHY) correction $E_{\text{\tiny LHY}}$ in Eq.~\eqref{2_comp_quantum_correction} arises due to the zero-point fluctuations of the Bogoliubov modes of the binary mixture, which for 1D is given by Eq.~\eqref{LHY_energy_correction}. For the two species having equal masses $m_{1}=m_{2}=m$, the integral in Eq.~\eqref{LHY_energy_correction} can be solved exactly,
\begin{align}
\frac{E_{\text{\tiny LHY}}}{\mathcal{E}_0} &= -\frac{L}{2\pi\xi}\frac{\sqrt{2}\chi^{-3/2}}{3}\sum_{\pm}\left(1+\chi^2\pm \sqrt{\left(1-\chi^2\right)^2+4\eta\chi^2} \right)^{3/2} \notag \\
%------
\Rightarrow E_{\text{\tiny LHY}} &= -\frac{L}{2\pi\hbar}\frac{\sqrt{2m}(g_{2}n_{2})^{3/2}}{3}\sum_{\pm}\left(1+\chi^2\pm \sqrt{\left(1-\chi^2\right)^2+4\eta\chi^2} \right)^{3/2} \label{LHY_analytical} 
\end{align}
\end{widetext}
where we have introduced the scales $\mathcal{E}_{0}=\sqrt{g_{1}n_{1}g_{2}n_{2}}=\hbar^2/\xi^2\sqrt{m_{1}m_{2}}$, $\xi^2=\hbar^2/\sqrt{m_{1}m_{2}g_{1}g_{2}n_{1}n_{2}}$, as well as the dimensionless parameters $\chi=\sqrt{g_{1}n_{1}/g_{2}n_{2}}$, $\eta = g_{12}^2/g_{1}g_{2}$. Note that, the LHY correction is attractive in 1D \cite{astrakharchik1,petrov2}. 

\subsection{Fr\"{o}hlich terms and quantum correction due to impurities}
\label{quantum_correction_due_to_impurities_appendix}
In this section, we derive the quantum correction arising due to the interaction between the impurities and the two-component Bose gas. By applying the transformations given in Eq.~\eqref{inv_bog_transformation}, the Fr\"ohlich part $\hat{\mathcal{H}}_{\rm F}$ can be written in terms of the Bogoliubov operators $\hat{\alpha}_{i,k}$  as follows,
\begin{align}
\hat{\mathcal{H}}_{\rm F} =& \sum_{i,k \neq 0} \lambda_{i}\sqrt{n_{i}}\, \hat{\rho}_{\text{\tiny I},k}(\hat{a}_{i,k}+\hat{a}^\dagger_{i,-k})\notag\\
=&\sum_{i,k \neq 0}\lambda_{i}\sqrt{n_i}\,\hat{\rho}_{\text{\tiny I},k}\bigg(\mathcal{N}^{(i)}_{+,k}\frac{\hbar^2k^2}{m_{i}}\hat{\alpha}_{1,k}+\mathcal{N}^{(i)}_{+,k}\frac{\hbar^2k^2}{m_i}\hat{\alpha}^\dagger_{1,-k}\notag\\
&+\mathcal{N}^{(i)}_{-,k}\frac{\hbar^2k^2}{m_{i}}\hat{\alpha}_{2,k}+\mathcal{N}^{(i)}_{-,k}\frac{\hbar^2k^2}{m_i}\hat{\alpha}^\dagger_{2,-k}\bigg)
\label{Frohlic_terms}
\end{align}
By using the second order perturbation theory for $\hat{\mathcal{H}}_{\rm F}$, we calculate the additional quantum fluctuation terms (other than the LHY term in Eq.~\eqref{LHY_energy_correction}) arising due to impurities. Here, we have assumed dilute gas of impurities i.e. $n_{\rm I} \ll n_{i}$, so that the perturbation theory is applicable. The second order perturbation theory states that,
\begin{eqnarray}
E^{(\rm I)} = -\sum_{\nu}\frac{|\langle \Omega | \hat{\mathcal{H}}_{\rm F}| \nu \rangle|^2}{(E_{\nu}-E_{0})}
\label{correction_terms} 
\end{eqnarray}
where $\ket{\Omega}$ is the Bogoliubov vacuum with energy $E_{0}$ and $\ket{\nu}\neq\ket{\Omega}$ are appropriately chosen excited states  with energy $E_{\nu}$. For Bogoliubov excitations of species $1$ and $2$, the matrix elements $\langle \Omega |\hat{\mathcal{H}}_{\rm F}| \nu \rangle$ are given by,
\begin{subequations}
\begin{align}
&\langle \Omega | \hat{\mathcal{H}}_{\rm F} | 1 \rangle \notag\\
&= \mathcal{N}^{(1)}_{+,k}\lambda_{1}\sqrt{n_{1}n_{\rm I}}\,\frac{\hbar^2 k^2}{m_{1}} + \mathcal{N}^{(2)}_{+,k}\lambda_{2}\sqrt{n_{2}n_{\rm I}}\,\frac{\hbar^2 k^2}{m_{2}} \label{matrix_element_species_1}\\
&\langle \Omega | \hat{\mathcal{H}}_{\rm F} | 2 \rangle \notag\\
&= \mathcal{N}^{(1)}_{-,k}\lambda_{1}\sqrt{n_{1}n_{\rm I}}\,\frac{\hbar^2 k^2}{m_{1}} + \mathcal{N}^{(2)}_{-,k}\lambda_{2}\sqrt{n_{2}n_{\rm I}}\,\frac{\hbar^2 k^2}{m_{2}} \label{matrix_element_species_2}
\end{align}
\end{subequations}
Substituting the above equations in Eq.~\eqref{correction_terms}, we obtain the following,
\begin{widetext}
\begin{subequations}
\begin{align}
\frac{|\langle \Omega | \hat{\mathcal{H}}_{\rm F} | 1 \rangle|^2}{(\mathcal{E}_{+,k}+\mathcal{E}_{{\rm I},k})}
&= \frac{\lambda^2_{1}\,n_{1}n_{\rm I}\,(\hbar^2k^2/m_{1})(\mathcal{E}^{2}_{+,k}-\mathcal{E}^{2}_{2,k})}{2\mathcal{E}_{+,k}(\mathcal{E}^{2}_{+,k}-\mathcal{E}^{2}_{-,k})(\mathcal{E}_{+,k}+\mathcal{E}_{{\rm I},k})}+\frac{\lambda^2_{2}\,n_{2}n_{\rm I}\,(\hbar^2k^2/m_{2})(\mathcal{E}^{2}_{+,k}-\mathcal{E}^{2}_{1,k})}{2\mathcal{E}_{+,k}(\mathcal{E}^{2}_{+,k}-\mathcal{E}^{2}_{-,k})(\mathcal{E}_{+,k}+\mathcal{E}_{{\rm I},k})}+\frac{g_{12}\,\lambda_{1}\lambda_{2}\,n_{1}n_{2}n_{\rm I}\,(\hbar^2k^2/\sqrt{m_{1}m_{2}})^2}{\mathcal{E}_{+,k}(\mathcal{E}^{2}_{+,k}-\mathcal{E}^{2}_{-,k})(\mathcal{E}_{+,k}+\mathcal{E}_{{\rm I},k})} \label{quantum_correction_species_1}\\
%-----------------------
\frac{|\langle \Omega | \hat{\mathcal{H}}_{\rm F} | 2 \rangle|^2}{(\mathcal{E}_{-,k}+\mathcal{E}_{{\rm I},k})} 
&= \frac{\lambda^2_{1}\,n_{1}n_{\rm I}\,(\hbar^2k^2/m_{1})(\mathcal{E}^{2}_{-,k}-\mathcal{E}^{2}_{2,k})}{2\mathcal{E}_{-,k}(\mathcal{E}^{2}_{-,k}-\mathcal{E}^{2}_{+,k})(\mathcal{E}_{-,k}+\mathcal{E}_{{\rm I},k})}+\frac{\lambda^2_{2}\,n_{2}n_{\rm I}\,(\hbar^2k^2/m_{2})(\mathcal{E}^{2}_{-,k}-\mathcal{E}^{2}_{1,k})}{2\mathcal{E}_{-,k}(\mathcal{E}^{2}_{-,k}-\mathcal{E}^{2}_{+,k})(\mathcal{E}_{-,k}+\mathcal{E}_{{\rm I},k})}+\frac{g_{12}\,\lambda_{1}\lambda_{2}\,n_{1}n_{2}n_{\rm I}\,(\hbar^2k^2/\sqrt{m_{1}m_{2}})^2}{\mathcal{E}_{-,k}(\mathcal{E}^{2}_{-,k}-\mathcal{E}^{2}_{+,k})(\mathcal{E}_{-,k}+\mathcal{E}_{{\rm I},k})}
\label{quantum_correction_species_2}
\end{align}
\end{subequations}
where we have used the relation $(\mathcal{E}^{2}_{\pm,k}-\mathcal{E}^{2}_{1,k})(\mathcal{E}^{2}_{\pm,k}-\mathcal{E}^{2}_{2,k}) = g^{2}_{12}n_{1}n_{2}(\hbar^2k^2/\sqrt{m_{1}m_{2}})^2$ in the third term of the above equations, and $\mathcal{E}_{{\rm I},k}=\hbar^2k^2/2m_{\rm I}$ denotes the excitation energy of the impurity state.
By adding the first terms of Eq.~\eqref{quantum_correction_species_1} and Eq.~\eqref{quantum_correction_species_2}, we obtain the following,
\begin{align}
E^{({\rm I})}_{1,k}=&-\frac{\lambda^2_{1}n_{1}n_{\rm I}}{2}\frac{\hbar^2k^2}{m_{1}}\left[\frac{(\mathcal{E}^{2}_{+,k}-\mathcal{E}^{2}_{2,k})\,\mathcal{E}^{-1}_{+,k}}{(\mathcal{E}^{2}_{+,k}-\mathcal{E}^{2}_{\mp,k})(\mathcal{E}_{+,k}+\mathcal{E}_{{\rm I},k})} + \frac{(\mathcal{E}^{2}_{-,k}-\mathcal{E}^{2}_{2,k})\,\mathcal{E}^{-1}_{-,k}}{(\mathcal{E}^{2}_{-,k}-\mathcal{E}^{2}_{+,k})(\mathcal{E}_{-,k}+\mathcal{E}_{{\rm I},k})} \right] \notag\\
%-----
=&-\frac{\lambda^2_{1}n_{1}n_{\rm I}}{2}\frac{\hbar^2k^2}{m_{1}}\left[\frac{(\mathcal{E}^{2}_{+,k}-\mathcal{E}^{2}_{2,k})(\mathcal{E}_{+,k}-\mathcal{E}_{{\rm I},k})\,\mathcal{E}^{-1}_{+,k}}{(\mathcal{E}^{2}_{+,k}-\mathcal{E}^{2}_{-,k})(\mathcal{E}^{2}_{+,k}-\mathcal{E}^{2}_{{\rm I},k})} + \frac{(\mathcal{E}^{2}_{-,k}-\mathcal{E}^{2}_{2,k})(\mathcal{E}_{-,k}-\mathcal{E}_{{\rm I},k})\,\mathcal{E}^{-1}_{-,k}}{(\mathcal{E}^{2}_{-,k}-\mathcal{E}^{2}_{+,k})(\mathcal{E}^{2}_{-,k}-\mathcal{E}^{2}_{{\rm I},k})}\right] \notag\\
%----
=&-\frac{\lambda^2_{1}n_{1}n_{\rm I}}{2}\frac{\hbar^2k^2}{m_{1}}\Bigg[\frac{(\mathcal{E}^{2}_{+,k}-\mathcal{E}^{2}_{2,k})}{(\mathcal{E}^{2}_{+,k}-\mathcal{E}^{2}_{-,k})(\mathcal{E}^{2}_{+,k}-\mathcal{E}^{2}_{{\rm I},k})} + \frac{(\mathcal{E}^{2}_{-,k}-\mathcal{E}^{2}_{2,k})}{(\mathcal{E}^{2}_{-,k}-\mathcal{E}^{2}_{+,k})(\mathcal{E}^{2}_{-,k}-\mathcal{E}^{2}_{{\rm I},k})} \notag\\
&- \frac{(\mathcal{E}^{2}_{+,k}-\mathcal{E}^{2}_{2,k})\,\mathcal{E}^{-1}_{+,k}\,\mathcal{E}_{{\rm I},k}}{(\mathcal{E}^{2}_{+,k}-\mathcal{E}^{2}_{-,k})(\mathcal{E}^{2}_{+,k}-\mathcal{E}^{2}_{{\rm I},k})}
- \frac{(\mathcal{E}^{2}_{-,k}-\mathcal{E}^{2}_{2,k})\,\mathcal{E}^{-1}_{-,k}\,\mathcal{E}_{{\rm I},k}}{(\mathcal{E}^{2}_{-,k}-\mathcal{E}^{2}_{+,k})(\mathcal{E}^{2}_{-,k}-\mathcal{E}^{2}_{{\rm I},k})} \Bigg] \notag\\
%----
=&-\frac{\lambda^2_{1}n_{1}n_{\rm I}}{2}\frac{\hbar^2k^2}{m_{1}}\Bigg[\frac{(\mathcal{E}^{2}_{+,k}-\mathcal{E}^{2}_{2,k})(\mathcal{E}^{2}_{-,k}-\mathcal{E}^{2}_{{\rm I},k})-(\mathcal{E}^{2}_{-,k}-\mathcal{E}^{2}_{2,k})(\mathcal{E}^{2}_{+,k}-\mathcal{E}^{2}_{{\rm I},k})}{(\mathcal{E}^{2}_{+,k}-\mathcal{E}^{2}_{-,k})(\mathcal{E}^{2}_{+,k}-\mathcal{E}^{2}_{{\rm I},k})(\mathcal{E}^{2}_{-,k}-\mathcal{E}^{2}_{{\rm I},k})} \notag\\
& -\frac{(\mathcal{E}^{2}_{+,k}-\mathcal{E}^{2}_{2,k})\,\mathcal{E}^{-1}_{+,k}\,\mathcal{E}_{{\rm I},k}}{(\mathcal{E}^{2}_{+,k}-\mathcal{E}^{2}_{-,k})(\mathcal{E}^{2}_{+,k}-\mathcal{E}^{2}_{{\rm I},k})} 
-\frac{(\mathcal{E}^{2}_{-,k}-\mathcal{E}^{2}_{2,k})\,\mathcal{E}^{-1}_{-,k}\,\mathcal{E}_{{\rm I},k}}{(\mathcal{E}^{2}_{-,k}-\mathcal{E}^{2}_{+,k})(\mathcal{E}^{2}_{-,k}-\mathcal{E}^{2}_{{\rm I},k})}\Bigg] \notag\\
%-----------
=&-\frac{\lambda^2_{1}n_{1}n_{\rm I}}{2}\frac{\hbar^2k^2}{m_{1}}\Bigg[ -\frac{(\mathcal{E}^{2}_{+,k}-\mathcal{E}^{2}_{-,k})(\mathcal{E}^{2}_{{\rm I},k}-\mathcal{E}^{2}_{2,k})}{(\mathcal{E}^{2}_{+,k}-\mathcal{E}^{2}_{-,k})(\mathcal{E}^{2}_{+,k}-\mathcal{E}^{2}_{{\rm I},k})(\mathcal{E}^{2}_{-,k}-\mathcal{E}^{2}_{{\rm I},k})} \notag\\
&-\frac{(\mathcal{E}^{2}_{+,k}-\mathcal{E}^{2}_{2,k})\,\mathcal{E}^{-1}_{+,k}\,\mathcal{E}_{{\rm I},k}}{(\mathcal{E}^{2}_{+,k}-\mathcal{E}^{2}_{-,k})(\mathcal{E}^{2}_{+,k}-\mathcal{E}^{2}_{{\rm I},k})}
 -\frac{(\mathcal{E}^{2}_{-,k}-\mathcal{E}^{2}_{2,k})\,\mathcal{E}^{-1}_{-,k}\,\mathcal{E}_{{\rm I},k}}{(\mathcal{E}^{2}_{-,k}-\mathcal{E}^{2}_{+,k})(\mathcal{E}^{2}_{-,k}-\mathcal{E}^{2}_{{\rm I},k})}  \Bigg] \notag\\
%------------
=&\,\frac{\lambda^2_{1}n_{1}n_{\rm I}}{2}\frac{\hbar^2k^2}{m_{1}}\,\mathcal{E}_{{\rm I},k}\,\Bigg[\frac{(\mathcal{E}^{2}_{{\rm I},k}-\mathcal{E}^{2}_{2,k})\,\mathcal{E}^{-1}_{{\rm I},k}}{(\mathcal{E}^{2}_{{\rm I},k}-\mathcal{E}^{2}_{+,k})(\mathcal{E}^{2}_{{\rm I},k}-\mathcal{E}^{2}_{-,k})}
+\frac{(\mathcal{E}^{2}_{+,k}-\mathcal{E}^{2}_{2,k})\,\mathcal{E}^{-1}_{+,k}}{(\mathcal{E}^{2}_{+,k}-\mathcal{E}^{2}_{-,k})(\mathcal{E}^{2}_{+,k}-\mathcal{E}^{2}_{{\rm I},k})} 
+\frac{(\mathcal{E}^{2}_{-,k}-\mathcal{E}^{2}_{2,k})\,\mathcal{E}^{-1}_{-,k}}{(\mathcal{E}^{2}_{-,k}-\mathcal{E}^{2}_{+,k})(\mathcal{E}^{2}_{-,k}-\mathcal{E}^{2}_{{\rm I},k})}
 \Bigg] \notag\\
%-----------
E^{({\rm I})}_{1,k}=&\,\frac{\lambda^2_{1}n_{1}n_{\rm I}}{4}\frac{\hbar^4k^4}{m_{1}m_{\rm I}}\Bigg[\frac{(\mathcal{E}^{2}_{{\rm I},k}-\mathcal{E}^{2}_{2,k})\,\mathcal{E}^{-1}_{{\rm I},k}}{(\mathcal{E}^{2}_{{\rm I},k}-\mathcal{E}^{2}_{+,k})(\mathcal{E}^{2}_{{\rm I},k}-\mathcal{E}^{2}_{-,k})}
+\frac{(\mathcal{E}^{2}_{+,k}-\mathcal{E}^{2}_{2,k})\,\mathcal{E}^{-1}_{+,k}}{(\mathcal{E}^{2}_{+,k}-\mathcal{E}^{2}_{-,k})(\mathcal{E}^{2}_{+,k}-\mathcal{E}^{2}_{{\rm I},k})} 
+\frac{(\mathcal{E}^{2}_{-,k}-\mathcal{E}^{2}_{2,k})\,\mathcal{E}^{-1}_{-,k}}{(\mathcal{E}^{2}_{-,k}-\mathcal{E}^{2}_{+,k})(\mathcal{E}^{2}_{-,k}-\mathcal{E}^{2}_{{\rm I},k})}
 \Bigg] \label{first_quantum_correction_discrete}
\end{align}
Similarly, by adding the second as well as third term of Eq.~\eqref{quantum_correction_species_1} and Eq.~\eqref{quantum_correction_species_2}, we obtain $E^{({\rm I})}_{2,k}$ and $E^{({\rm I})}_{12,k}$, respectively,
\begin{align}
E^{({\rm I})}_{2,k}=&\,\frac{\lambda^2_{2}n_{2}n_{\rm I}}{4}\frac{\hbar^4k^4}{m_{2}m_{\rm I}}\Bigg[\frac{(\mathcal{E}^{2}_{{\rm I},k}-\mathcal{E}^{1}_{1,k})\,\mathcal{E}^{-1}_{{\rm I},k}}{(\mathcal{E}^{2}_{{\rm I},k}-\mathcal{E}^{2}_{+,k})(\mathcal{E}^{2}_{{\rm I},k}-\mathcal{E}^{2}_{-,k})}
+\frac{(\mathcal{E}^{2}_{+,k}-\mathcal{E}^{2}_{1,k})\,\mathcal{E}^{-1}_{+,k}}{(\mathcal{E}^{2}_{+,k}-\mathcal{E}^{2}_{-,k})(\mathcal{E}^{2}_{+,k}-\mathcal{E}^{2}_{{\rm I},k})} 
+\frac{(\mathcal{E}^{2}_{-,k}-\mathcal{E}^{2}_{1,k})\,\mathcal{E}^{-1}_{-,k}}{(\mathcal{E}^{2}_{-,k}-\mathcal{E}^{2}_{+,k})(\mathcal{E}^{2}_{-,k}-\mathcal{E}^{2}_{{\rm I},k})}\Bigg] \label{second_quantum_correction_discrete}\\
%---------
E^{({\rm I})}_{12,k}=&\,\frac{\lambda_{1}\lambda_{2}n_{1}n_{2}n_{\rm I}}{2}\frac{g_{12}\,\hbar^6k^6}{m_{1}m_{2}m_{\rm I}}\Bigg[\frac{\mathcal{E}^{-1}_{+,k}}{(\mathcal{E}^{2}_{+,k}-\mathcal{E}^{2}_{-,k})(\mathcal{E}^{2}_{+,k}-\mathcal{E}^{2}_{{\rm I},k})} 
+\frac{\mathcal{E}^{-1}_{-,k}}{(\mathcal{E}^{2}_{-,k}-\mathcal{E}^{2}_{+,k})(\mathcal{E}^{2}_{-,k}-\mathcal{E}^{2}_{{\rm I},k})}  
+\frac{\mathcal{E}^{-1}_{{\rm I},k}}{(\mathcal{E}^{2}_{{\rm I},k}-\mathcal{E}^{2}_{+,k})(\mathcal{E}^{2}_{{\rm I},k}-\mathcal{E}^{2}_{-,k})} \Bigg] \label{third_quantum_correction_discrete}
\end{align}
Therefore, the quantum correction in energy arising due to the interaction with the mobile impurities is given by,
\begin{eqnarray}
E^{(\rm I)} = \sum_{k} E^{(\rm I)}_{1,k} + E^{(\rm I)}_{2,k} + E^{(\rm I)}_{12,k}
\label{quantum_correction_impurity}
\end{eqnarray}
We can now write the above equation Eq.~\eqref{quantum_correction_impurity} in the integral form, which can be solved exactly for equal masses of the bosonic species $m_{1}=m_{2}=m$,
\begin{align}
\frac{E^{(\rm I)}}{\mathcal{E}_0}&=L\int^{\infty}_{-\infty} \frac{dk}{2\pi}\frac{\left(E^{(\rm I)}_{1,k}+E^{(\rm I)}_{2,k}+E^{(\rm I)}_{12,k}\right)}{\mathcal{E}_{0}}\notag\\
&=\frac{L}{2\pi\xi}\sqrt{2}\chi^{1/2}\Lambda(\alpha)\!\left(\left(1-\chi^2\right)^2\!+\!4\eta\chi^2 \right)^{-1/2}\!\!\left(\frac{\lambda^2_{1}}{g_{1}g_{2}}\frac{n_{\rm I}}{n_{2}}\mathcal{F}(\chi,\eta)\!+\!\frac{\lambda^2_{2}}{g_{1}g_{2}}\frac{n_{\rm I}}{n_{1}}\mathcal{G}(\chi,\eta)\!+\!\frac{\lambda_{1}\lambda_{2}}{g_{1}g_{2}}\frac{4\sqrt{\eta}\,n_{\rm I}}{\sqrt{n_{1}n_{2}}}\mathcal{K}(\chi,\eta)\right)\\
\Rightarrow E^{(\rm I)} &= \frac{L}{2\pi\hbar}g_{1}n_{1}[g_{2}n_{2}]^{1/2}\sqrt{2m} \Lambda(\alpha)\!\left(\left(1-\chi^2\right)^2\!+\!4\eta\chi^2 \right)^{-1/2}\!\left(\frac{\lambda^2_{1}}{g_{1}g_{2}}\frac{n_{\rm I}}{n_{2}}\mathcal{F}(\chi,\eta)\!+\!\frac{\lambda^2_{2}}{g_{1}g_{2}}\frac{n_{\rm I}}{n_{1}}\mathcal{G}(\chi,\eta)\!+\!\frac{\lambda_{1}\lambda_{2}}{g_{1}g_{2}}\frac{4\sqrt{\eta}\,n_{\rm I}}{\sqrt{n_{1}n_{2}}}\mathcal{K}(\chi,\eta)\right) \notag
\end{align}
where $\Lambda(\alpha) = \sin^{-1}({\sqrt{1-\alpha^2}})/\sqrt{1-\alpha^2}$, $\alpha = m/m_{\rm I}$, and the functions $\mathcal{F}(\chi,\eta)$, $\mathcal{G}(\chi,\eta)$, $\mathcal{K}(\chi,\eta)$ are given by,
\begin{subequations}
\begin{align}
&\mathcal{F}(\chi,\eta)=\frac{\left(1-\chi^2-\sqrt{\left(1-\chi^2\right)^2+4\eta\chi^2}\,\right)}{\left(1+\chi^2+\sqrt{\left(1-\chi^2\,\right)^2+4\eta\chi^2}\,\right)^{1/2}}-\frac{\left(1-\chi^2+\sqrt{\left(1-\chi^2\right)^2+4\eta\chi^2}\,\right)}{\left(1+\chi^2-\sqrt{\left(1-\chi^2\,\right)^2+4\eta\chi^2}\,\right)^{1/2}}  \\
&\mathcal{G}(\chi,\eta)=\frac{\left(1-\chi^2-\sqrt{\left(1-\chi^2\right)^2+4\eta\chi^2}\,\right)}{\left(1+\chi^2-\sqrt{\left(1-\chi^2\,\right)^2+4\eta\chi^2}\,\right)^{1/2}}-\frac{\left(1-\chi^2+\sqrt{\left(1-\chi^2\,\right)^2+4\eta\chi^2}\,\right)}{\left(1+\chi^2+\sqrt{\left(1-\chi^2\right)^2+4\eta\chi^2}\,\right)^{1/2}}  \\
&\mathcal{K}(\chi,\eta)=\left[\left(1+\chi^2-\sqrt{\left(1-\chi^2\,\right)^2+4\eta\chi^2}\,\right)^{\!-1/2}-\left(1+\chi^2+\sqrt{\left(1-\chi^2\right)^2+4\eta\chi^2}\,\right)^{\!-1/2}\right]\chi
\end{align}
\end{subequations}
\end{widetext}
Consequently, the total energy functional computed from Eq.~\eqref{background_expansion} for the homogeneous case is given by,
\begin{align}
&E[n_{1},n_{2},n_{\rm I}]& \notag\\
&=L(g_{1}n^{2}_{1}/2+g_{2}n^{2}_{2}/2+g_{12}n_{1}n_{2}) + L(\lambda_{1}n_{\rm I}n_{1}+\lambda_{2}n_{\rm I}n_{2})\notag\\
&\quad\,+E_{\text{\tiny LHY}}[n_{1},n_{2}] + E^{(\rm I)}[n_{1},n_{2},n_{\rm I}]
\label{energy_density_homogeneous}
\end{align}

\section{Local density approximation}
\label{appendix_local_density_approx}
Next, we write the above mentioned energy functional in Eq.~\eqref{energy_density_homogeneous} within the local density approximation (LDA) with $\psi_{\rm I}(x)$ being the impurity wavefunction and $\phi_{1(2)}(x)$ being the wavefunction of the two components of the binary mixture,
%\begin{eqnarray}
%E[\phi_{1}(x),\phi_{2}(x),\psi_{\rm I}(x)] &\!\!=\!\!& E_{\rm BB}[\phi_{1}(x),\phi_{2}(x)]+E_{\rm I}[\psi_{\rm I}(x)] \notag\\
%&\!\!+\!\!&E_{\rm IB}[\psi_{\rm I}(x),\phi_{1}(x),\phi_{2}(x)]
%\label{impurity_boson_energy_functional_LDA}
%\end{eqnarray}
\begin{widetext}
\begin{align}
E[\phi_{1}(x),\phi_{2}(x),\psi_{\rm I}(x)] = E_{\rm BB}[\phi_{1}(x),\phi_{2}(x)]+E_{\rm I}[\psi_{\rm I}(x)] + E_{\rm IB}[\psi_{\rm I}(x),\phi_{1}(x),\phi_{2}(x)]
\label{impurity_boson_energy_functional_LDA}
\end{align}

where the individual terms are given by the following,
\begin{itemize}
\begin{subequations}
\item Energy functional of the two-component Bose gas:
\begin{align}
E_{\rm BB}[\phi_{1}(x),\phi_{2}(x)]
&= \sum_{i=1,2} \left[\int\!dx\,\phi_{i}^{\star}(x)\left(-\frac{\hbar^2}{2m}\frac{d^2}{dx^2}\right)\phi_{i}(x)+\int\!dx\,\frac{g_{i}n^2_{i}(x)}{2} \right]+ \int\!dx\, g_{12}n_{1}(x)n_{2}(x) \notag\\
&\quad\,-\int\!dx\,\frac{\sqrt{2m}[g_{2}n_{2}(x)]^{3/2}}{2\pi\hbar}\sum_{\pm}\frac{1}{3}\Bigg(1+\chi^2(x)\pm\sqrt{[1-\chi^2(x)]^2+4\eta\chi^2(x)}\Bigg)^{\!3/2}
\label{boson_energy_functional_LDA}
\end{align}
%------------------------------
\item Energy functional of the impurities:
\begin{align}
E_{\rm I}[\psi_{\rm I}(x)] &=  \int\!dx\, \psi^{\star}_{\rm I}(x)\left(-\frac{\hbar^2}{2m_{\rm I}}\frac{d^2}{dx^2}\right)\psi_{\rm I}(x) 
\label{impurity_energy_functional_LDA}
\end{align}
%-------------------------------
\item Energy functional of the interaction between the impurities and binary mixture:
\begin{align}
E_{\rm IB}[\psi_{\rm I}(x),\phi_{1}(x),\phi_{2}(x)]
&= \sum_{i=1,2} \int\!dx\, \lambda_{i}n_{i}(x)n_{\rm I}(x) 
+\int\!dx\,\frac{g_{1}n_{1}(x)[g_{2}n_{2}(x)]^{1/2}\sqrt{2m}}{2\pi \hbar} \Lambda(\alpha)\Bigg(\left[1-\chi^2(x)\right]^2+4\eta\chi^2(x) \Bigg)^{\!-1/2}\notag\\
&\quad\, \left(\frac{\lambda^2_{1}}{g_{1}g_{2}}\frac{n_{\rm I}(x)}{n_{2}(x)}\mathcal{F}[\chi(x),\eta]+\frac{\lambda^2_{2}}{g_{1}g_{2}}\frac{n_{\rm I}(x)}{n_{1}(x)}\mathcal{G}[\chi(x),\eta]+\frac{\lambda_{1}\lambda_{2}}{g_{1}g_{2}}\frac{4\sqrt{\eta}\,n_{\rm I}(x)}{\sqrt{n_{1}(x)n_{2}(x)}}\mathcal{K}[\chi(x),\eta]\right)
\end{align}
\end{subequations}
\end{itemize}
with $n_{i}(x)=\phi^{\star}(x)\phi(x)$, $n_{\rm I}(x) = \psi^{\star}_{\rm I}(x)\psi_{\rm I}(x)$, and $\chi(x) = \sqrt{g_{1}n_{1}(x)/g_{2}n_{2}(x)}$.

\subsection{Equations of motion for two-component Bose gas}
The coupled Gross-Pitaevskii equations obtained from the Energy functional in Eq.~\eqref{impurity_boson_energy_functional_LDA} are given by,
\begin{subequations}
\begin{eqnarray}
\imath \hbar\frac{\partial}{\partial t}\phi_{1}(x,t) &=& \left(\frac{-\hbar^2}{2m}\frac{\partial^2 }{\partial x^2} + g_{1}n_{1}(x,t) + g_{12}n_{2}(x,t)+\lambda_{1}n_{\rm I}(x,t) + \Delta\mu^{(1)}_{\text{\tiny LHY}}(x,t)\right)\phi_{1}(x,t) \\
\imath \hbar\frac{\partial}{\partial t}\phi_{2}(x,t) &=& \left(\frac{-\hbar^2}{2m}\frac{\partial^2 }{\partial x^2} + g_{2}n_{2}(x,t) + g_{12}n_{1}(x,t)+\lambda_{2}n_{\rm I}(x,t) + \Delta\mu^{(2)}_{\text{\tiny LHY}}(x,t)\right)\phi_{2}(x,t) \\
\imath \hbar\frac{\partial}{\partial t}\psi_{\rm I}(x,t) &=& \left(\frac{-\hbar^2}{2m_{\rm I}}\frac{\partial^2 }{\partial x^2} +\lambda_{1}n_{1}(x,t) + \lambda_{2}n_{2}(x,t) + \Delta\mu_{\rm I}(x,t)\right)\psi_{\rm I}(x,t)
\end{eqnarray}
\label{coupled_GPE}
\end{subequations}
where the correction to the chemical potentials are,
\begin{subequations}
\begin{align}
\Delta \mu^{(1)}_{\text{\tiny LHY}}(x) &= -\frac{g_{1}[g_{2}n_{2}(x)]^{1/2}\sqrt{2m}}{4\pi\hbar} \sum_{\pm} \bigg(1+\chi^2(x)\pm\sqrt{[1-\chi^2(x)]^2+4\eta\chi^2(x)} \bigg)^{1/2}\Bigg( 1 \pm \frac{\left(2\eta+\chi^2(x)-1\right)}{\sqrt{[1-\chi^2(x)]^2+4\eta\chi^2(x)}}\Bigg) \\
%--------------------
\Delta \mu^{(2)}_{\text{\tiny LHY}}(x) &= -\frac{g_{2}[g_{2}n_{2}(x)]^{1/2}\sqrt{2m}}{4\pi\hbar} \sum_{\pm} \bigg(1+\chi^2(x)\pm\sqrt{[1-\chi^2(x)]^2+4\eta\chi^2(x)} \bigg)^{1/2}\Bigg( 1 \pm \frac{\left(2\eta\chi^2(x)-\chi^2(x)+1\right)}{\sqrt{[1-\chi^2(x)]^2+4\eta\chi^2(x)}}\Bigg) \\
%----------------------
\Delta \mu_{\rm I}(x) &= \frac{g_{1}n_{1}(x)[g_{2}n_{2}(x)]^{\frac{1}{2}}\sqrt{2m}}{2\pi \hbar} \frac{\sin^{-1}{(\sqrt{1-\alpha^2})}}{\sqrt{1-\alpha^2}}\!\Bigg(\left[1-\chi^2(x)\right]^2+4\eta\chi^2(x) \Bigg)^{\!-1/2}\notag\\
&\quad\,\left(\frac{\lambda^2_{1}}{g_{1}g_{2}}\frac{1}{n_{2}(x)}\mathcal{F}[\chi(x),\eta]\!+\!\frac{\lambda^2_{2}}{g_{1}g_{2}}\frac{1}{n_{1}(x)}\mathcal{G}[\chi(x),\eta]\!+\!\frac{\lambda_{1}\lambda_{2}}{g_{1}g_{2}}\frac{4\sqrt{\eta}}{\sqrt{n_{1}(x)n_{2}(x)}}\mathcal{K}[\chi(x),\eta]\right)
\end{align}
\end{subequations}
\end{widetext}
We point out that, the higher order correction in the chemical potential arising due to impurity atoms $\Delta \mu^{(i)}_{\rm I} = \partial E^{(\rm I)}/\partial n_{i} \propto n_{\rm I}$, and as a result, $\Delta \mu^{(i)}_{\rm I} \ll \Delta \mu_{\text{\tiny LHY}}$ can be neglected.
In the main text, we have considered  condensates with same density profile $n_{1}(x)=n_{2}(x)=n(x)$ and equal intra-species interaction $g_{1}=g_{2}=g$, for which we obtain an effective single component system with density $n(x)$ \cite{astrakharchik1,petrov2}. In that case, the above mentioned equations in Eq.~\eqref{coupled_GPE} reduce to Eq.~\eqref{effective_GPE}, which describe the time evolution of the effective single component condensate and the impurities.

\end{document}